\documentclass[12pt,epsf]{article}

\usepackage{graphicx}
\usepackage{epsfig}
\usepackage{amsmath}
\usepackage{multirow}
\usepackage{amssymb}
\usepackage{setspace}
\makeatletter
\let\@fnsymbol\@arabic
\makeatother

\begin{document}
\title{Indirect transmission and the effect of seasonal pathogen inactivation on infectious disease periodicity}
\author{Marguerite Robinson$^{a,*,}$\thanks{Present address: Institut Catal\`{a} de Ci\`{e}ncies del Clima (IC$^3$), C/ Doctor Trueta 203, 08005 Barcelona, Spain.}, Yannis Drossinos$^{a}$ \& Nikolaos I. Stilianakis$^{a,b}$.
 \\
\selectfont{\small{$^{a}$Joint Research Centre, European Commission, I-21027 Ispra (VA),
Italy}}
\\
\selectfont{\small{
$^{b}$Department of Biometry and Epidemiology, University of Erlangen-Nuremberg, Erlangen, Germany}}
\\
\selectfont{\small{$^*$Corresponding author, email: marguerite.robinson@ic3.cat, phone: +34 935679977}}}
\date{}
\maketitle


\begin{abstract}
The annual occurrence of many infectious diseases remains a constant
burden to public health systems. The seasonal patterns in
respiratory disease incidence observed in temperate regions have been
attributed to the impact of environmental conditions on pathogen
survival. A model describing the transmission of an infectious
disease by means of a pathogenic state capable of surviving in an
environmental reservoir outside of its host organism is presented in
this paper. The ratio of pathogen lifespan to the duration of the
infectious disease state is found to be a critical parameter in
determining disease dynamics. The introduction of a seasonally
forced pathogen inactivation rate identifies a time delay between
peak pathogen survival and peak disease incidence. The delay is
dependent on specific disease parameters and, for influenza,
decreases with increasing reproduction number. The observed seasonal
oscillations are found to have a period identical to that of the
seasonally forced inactivation rate and which is independent of the
duration of infection acquired immunity.\\

\end{abstract}


\section{Introduction}
\label{intro} Many diseases exhibit seasonal cycles in incidence
data, most notably childhood diseases such as measles and rubella (Anderson and May, 1991, Dowell, 2001).
For such diseases, it is widely accepted
that the observed cycles are linked to the timing of school terms
and the subsequent increase in contact rates among the
immunologically naive child population (Keeling and Rohani, 2001, Stone et al. 2007). The
mechanism driving the seasonal occurrence of influenza and other
respiratory infections is less well understood. While increased
contact rates due to indoor crowding during winter undoubtedly
facilitate greater disease transmission, it is more likely a
contributing factor and not the driving mechanism (Lofgren et al. 2007).
Seasonal influenza places a considerable burden on public health
systems with annual global incidence in the range 5 - 15\% of the
population, resulting in up to 500,000 deaths (Stohr, 2002).
Targeted interventions to reduce this burden could be significantly
improved if the seasonal
stimulus was better understood.\\

The cause of seasonality must be attributed to annual changes
associated with either the host or the infectious disease pathogen
(Grassly and Fraser, 2006). In addition to increased contact rates during
winter, higher transmission rates have been linked to changes in the
human immune system. Possible
variations in human immune function have been attributed to
fluctuations in melatonin secretion regulated by the annual
light-dark cycles and to deficiencies in vitamin D during the winter
months (Cannell et al. 2008, Lofgren et al. 2007, Nelson and Demas, 1996). Experiments on mice in an environmentally
controlled environment found that mice were more susceptible to
infection during the winter months (Schulman and Kilbourne, 1963). For diseases transmitted
through indirect pathways (e.g. respiratory droplets, fomites,
fungal spores, waterborne pathogens) the ability of the pathogen to
survive outside of its host must play a vital role in the
transmission process (Grassly and Fraser, 2006). This survival potential would
be greatly influenced by environmental factors. Early studies on the
survival of the influenza virus in air indicated that pathogen
survival peaks at low relative humidity and low temperatures
resulting in increased viral transmission (Harper, 1961, Hemmes et al. 1960).
Experiments to study the airborne transmission of influenza among
Guinea pigs further supported this theory indicating that
transmission among subjects peaked in cold dry air (Lowen et al. 2007).
However, while indoor relative humidity is minimized in winter,
outdoor humidity peaks. More recent studies identify absolute
humidity as a more likely seasonal driver, as both indoor and
outdoor absolute humidity display cyclic behavior in temperate
regions that minimizes in winter (Shaman and Kohn, 2009). Influenza virus
survival increases with low absolute humidity leading to increased
transmission in winter (Shaman et al. 2010). Other theories for
increased viral inactivation in summer consider ultraviolet
radiation and air toxicity, however, little work has been undertaken
to support such theories (Weber and Stilianakis, 2008).\\

A crucial factor to consider when assessing the impact of
environmental factors on seasonal transmission is the mode of
disease transmission. For respiratory diseases three modes of
pathogen transmission have been identified: droplet, contact and
airborne transmission (Weber and Stilianakis, 2008). Droplet transmission occurs
when, following an expiratory event by an infected individual, large
pathogen-carrying droplets (diameter $\gtrsim10\ \mu$m) are
deposited directly onto the mucous membranes of a susceptible
person. The direct nature of such a transmission pathway precludes a
significant impact by environmental factors on either the droplet
itself or its pathogen load. Conversely, the contact (through
fomites or direct human-to-human contact) and airborne (small
aerosol droplets with diameter $\lesssim10\ \mu$m) modes of
transmission render the pathogen vulnerable to environmental
conditions during prolonged periods spent external to its host
organism. For the specific case of the influenza virus, the
efficiency of contact transmission is determined by the survival
rate of the pathogen on  solid surfaces and human skin. The high
inactivation rates observed on hands can limit the occurrence of
transmission via the contact route, however, a continuous supply of
fomites from infected individuals could possibly counteract this
(Weber and Stilianakis, 2008). This mode of transmission is greatly influenced by
human behavior (e.g. cough etiquette, hand washing) and thus a
seasonal variation in its impact would be difficult to quantify.
Airborne droplets remain exposed to the ambient environmental
conditions for prolonged periods of time and could act as an
important agent driving seasonal disease incidence. For respiratory
diseases mediated by airborne droplets and for a spatially homogeneous distribution of susceptibles, pathogen removal is the
result of three distinct processes: gravitational settling, pathogen
inactivation and inhalation (Stilianakis and Drossinos, 2010). Environmental
factors are capable of influencing the first two of these processes
(Shaman and Kohn, 2009). The gravitational settling rate of a droplet is
determined from its diameter, which is influenced by evaporation
effects determined by the ambient air properties. However, a review
of experimental studies by Shaman and Kohn (2009) found that there
is insufficient evidence to support such a hypothesis. They
concluded that seasonal variations in environmental
conditions are most likely incorporated through the pathogen inactivation term.\\

The seasonality of infectious diseases is usually incorporated in
deterministic models through a time dependent transmission rate.
Outbreaks are typically simulated using a sinusoidal function or, in
the case of childhood diseases such as measles, a step function is
employed to represent school terms (Grassly and Fraser, 2006, Keeling and Rohani, 2001). Models
incorporating natural birth and death processes display damped
oscillations towards an endemic disease state (Fisman, 2007).
However, if the period of the applied seasonal forcing is close to
the intrinsic period of the damped oscillations then the two effects
can resonate to produce large amplitude seasonal oscillations
(Dushoff et al. 2004). An important consideration in any analysis of
seasonal influenza is the process of antigenic drift, whereby
continuous small changes in the virus requires the production of a
new vaccine each year. This phenomenon can be integrated into the
standard deterministic models by allowing recovered individuals to
lose their acquired immunity after a specified period of time
(Dushoff et al. 2004). An important factor neglected in the standard
deterministic models is the mechanism driving the seasonal
variation. If the theory that the impact of annual variations in
environmental conditions on pathogen survival/inactivation is indeed
valid then models should incorporate this phenomenon. In this paper, we present such a model.\\

In Section \ref{Sec2} we analyze a model for the spread of an
infectious disease by means of an intermediate free-living
pathogenic state, which is exposed to the ambient environmental
conditions. This general model, applicable to a variety of free-living organisms (e.g. viruses, bacteria, fungi, protozoa), was introduced by Anderson and May (1981) to describe indirect disease transmission
between free-living microparasites and their invertebrate hosts. Variations of the
basic model have been employed to describe the spread of a pathogen through a generalized
environmental state (Li et al. 2009), the spread of a waterborne bacteria (Tien and
Earn, 2010), bacterial and prion disease in livestock and wildlife (Nieuwhof 2009, Miller 2006), and the
transmission of respiratory diseases by airborne droplets (Stilianakis and Drossinos, 2010). The model
has even been adapted to describe the spread of fungal spores in a vineyard (Burie, 2006) and
the point release of an infectious agent (Reluga, 2004). With such a wide range of applications and descriptive abilities the intrinsic characteristics of the infectious agent are a vital model component and can radically impact the dynamics. A pathogen characteristic of primary importance is its ability to survive for prolonged periods outside of its host, which is directly influenced by environmental conditions and, thus, cannot be ignored in the quest to identify seasonal disease drivers. Therefore, we first analyse the general model and consider its behavior for both short-lived and long-lived pathogenic states. An important outcome of this analysis is that, in the case of short-lived pathogens, a quasi-steady state exists whereby the pathogen dynamics can be described in terms of the infected population alone. This quasi-steady approximation can also be applied to an autonomous model describing seasonal outbreaks. This model is analysed in Section \ref{Sec3}, where we consider the consequences of a seasonal variation in
the pathogen inactivation rate and how it impacts disease incidence.

\section{A model for indirect transmission of an infectious disease}
\label{Sec2} Many infectious diseases are primarily transmitted by
means of an intermediate environmental reservoir, in which
free-living pathogens are capable of surviving outside of the host
organism. For example, waterborne outbreaks can persist through the
shedding of pathogens by an infected individual into a water source
which is then ingested by a susceptible individual. Similarly,
respiratory diseases can be transmitted from person to person via
pathogen loaded airborne droplets or fomites expelled into the
environment by infected individuals during expiratory events (e.g.
coughing or sneezing). The efficiency of such transmission pathways
will depend on the ability of the pathogen to survive in the
intermediate reservoir (e.g. air, water). In this section we present
a generalized model for the transmission of an infection
by means of such a pathogen reservoir.\\

\begin{figure*}
\centering
\includegraphics[width=0.7\textwidth]{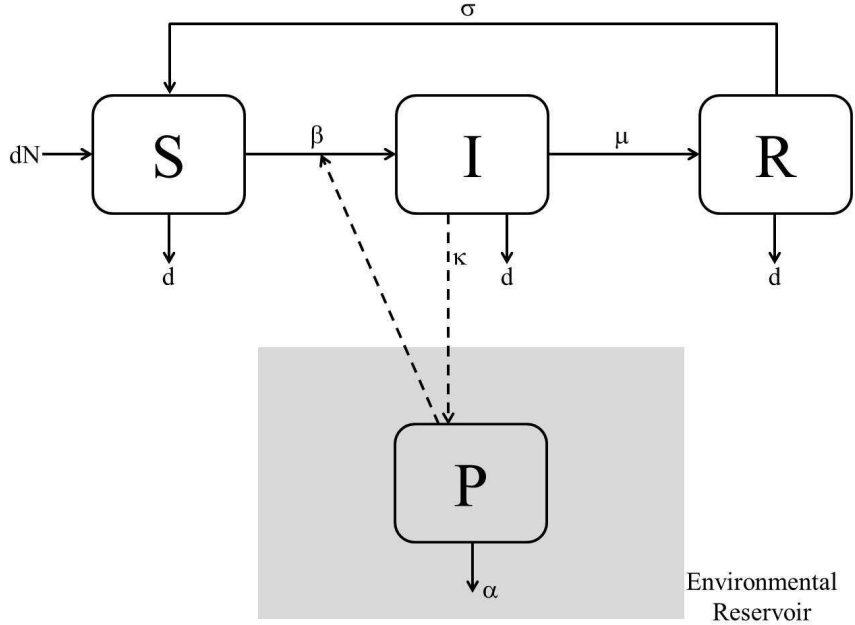}
\caption{Infection pathway for the transmission of a disease by a
free-living pathogen $P$ in a population of susceptible $S$,
infected $I$ and recovered $R$ individuals.} \label{fig:pathway}
\end{figure*}

Consider a closed population of $N$ individuals, of which $S(t)$ are
susceptible, $I(t)$  are infected and $R(t)$ are recovered. Infected
individuals shed pathogens into an intermediate reservoir, which are
free-living (outside of the host organism) and can transmit
infection through contact with the susceptible population, Figure
\ref{fig:pathway}. The total number of pathogens in the reservoir is
$P(t)$. A deterministic model for the disease dynamics is
\begin{align*}
\frac{dS}{dt}&=d(N-S)+\sigma(N-S-I)-\frac{\beta}{N}PS,\\
\frac{dI}{dt}&=\frac{\beta}{N}PS-(\mu+d)I,\\
\frac{dP}{dt}&=\kappa I-\alpha P,
\end{align*}
where the total human population $N=S+I+R$ is constant. The duration
of time spent in the infected state is $T_I=1/(\mu+d)$, where
$1/\mu$ is the average infectious disease period and
$1/d$ is a typical human lifespan. Infected individuals
generate pathogens, with a lifespan of $T_P=1/\alpha$, at a
rate $\kappa$. Infection acquired immunity is lost after a time
period of $1/\sigma$, with $\sigma=0$ if immunity is
permanent. The transmission rate per pathogen is $\beta$, determined
from the contact rate of a susceptible with a pathogen and the
probability of such a contact causing infection. The equation
describing pathogen dynamics assumes that pathogen numbers decrease
exponentially for positive $\alpha$. This could be modified to
describe pathogens with more complicated life cycles. The pathogen
removal rate $\alpha$ can be a complicated function of environmental
factors, depending on the
specific pathogen type and reservoir characteristics, but for now we assume it is constant.\\

The general model presented here is applicable to a wide variety of diseases such as cholera and influenza in humans (Tien and Earn 2010, Stilianakis and Drossinos 2010) and footrot and chronic wasting disease in animals (Nieuwhof 2009, Miller 2006). In the case of waterborne pathogens, this model has been expanded, specifically in the case of cholera, to include dose dependent infection rates (Code\c{c}o 2001, Bertuzzo 2010) and a minimum infectious dose (Joh, 2009). To analyse the general framework of the basic model we omit such disease dependent complications, but will consider this issue in relation to influenza in Section 3.3.\\

The model can be written in dimensionless form by scaling
\begin{equation}\label{scales}
S,I\sim N,\quad P\sim\frac{\kappa}{\alpha}N,\quad
t\sim\frac{1}{\mu+d},
\end{equation}
and the dimensionless system is
\begin{align}
\frac{dS}{dt}&=\phi(1-S)+\psi(1-S-I)-R_{0}PS,\label{eqn1}\\
\frac{dI}{dt}&=R_{0}PS-I,\label{eqn2}\\
\rho\frac{dP}{dt}&=I-P,\label{Feqn}
\end{align}
where dimensionless numbers are defined as
\begin{displaymath}
R_{0}=\frac{\beta\kappa}{\alpha(\mu+d)},\quad\quad\phi=\frac{d}{\mu+d},\quad\quad\psi=\frac{\sigma}{\mu+d},\quad\quad\rho=\frac{\mu+d}{\alpha}=\frac{T_P}{T_I}.
\end{displaymath}
The dimensionless number $\rho$ represents the ratio of the pathogen
lifespan to that of an infected individual. The dynamics of the
pathogen, satisfying (\ref{Feqn}), will be determined by the
magnitude of $\rho$. The magnitude of $R_0$ represents the average
number of secondary infections from a primary infection and is
typically $O(1)$. Finally, $\phi$ and $\psi$ determine how quickly
the pool of susceptible individuals is repopulated from natural
birth or death processes and loss of acquired immunity respectively.
Suitable initial conditions are
\begin{equation}\label{ics}
S(0)=S_0,\quad\quad I(0)=1-S_0=I_0,\quad\quad P(0)=P_0,
\end{equation}
where it is assumed that no pre-existing immunity is present in the population, $R(0)=0$.\\

The system (\ref{eqn1})-(\ref{Feqn}) has two equilibrium states. A
disease-free state $E_0=(1,0,0)$ and an endemic state $E_e$ where
\begin{equation}\label{Pe}
E_e=(S_e,I_e,P_e)=\left(\frac{1}{R_0},\frac{(\phi+\psi)(R_0-1)}{R_0(\psi+1)},\frac{(\phi+\psi)(R_0-1)}{R_0(\psi+1)}\right),
\end{equation}
which exists when at least one of $\phi$ or $\psi$ is nonzero and
$R_0>1$. It can be easily shown by a linear analysis that the
disease-free state is locally stable when $R_0<1$ and the initial
infection (\ref{ics}) dies out. Conversely, the endemic state is
locally stable when $R_0>1$ and the initial infection will spread
approaching the equilibrium state $E_e$ as $t\to\infty$. The
equilibrium point is a stable focus resulting in damped oscillations
towards $E_e$. As we are interested in the long-term seasonal
variation in the endemic
state we will assume that $R_0>1$ for our purposes.\\

\subsection{Short-lived pathogenic state}
\label{Sec 2.1} If the duration of the infectious disease state is
significantly longer than the lifespan of the pathogen, $T_P\ll
T_I$, then $\rho\ll1$. This implies that equation (\ref{Feqn})
rapidly achieves an equilibrium compared with the rest of the
system.
 We can assume that the left-hand side of equation (\ref{Feqn}) is approximately zero, $\rho\frac{dP}{dt}=0$, and employ the
 approximation
\begin{equation}\label{approx}
P\approx I.
\end{equation}
The system then reduces to the standard SIRS model
\begin{align}
\frac{dS}{dt}&=\phi(1-S)+\psi(1-S-I)-R_{0}IS,\label{SIRS1}\\
\frac{dI}{dt}&=(R_{0}S-1)I\label{SIRS2},
\end{align}
with initial conditions
\begin{equation}\label{ics2}
S(0)=S_0,\quad\quad I(0)=I_0,\quad\quad P(0)=I_0.
\end{equation}
\begin{figure*}
\includegraphics[width=1\textwidth]{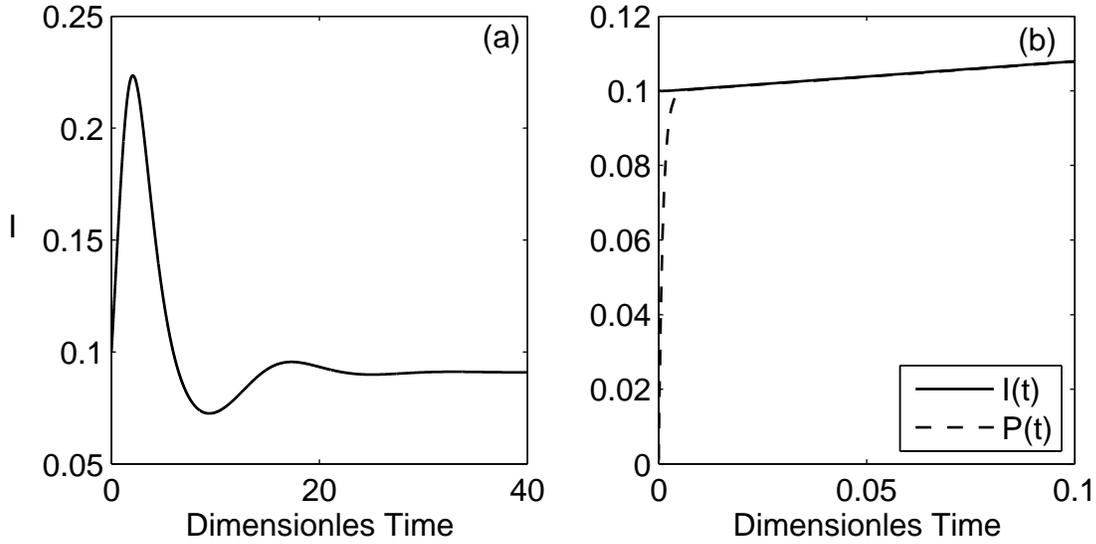}
\caption{Numerical solution of the SIRS system
(\ref{SIRS1})-(\ref{SIRS2}) with parameter values $\phi=\psi=0.1$,
$R_0=2$ and $\rho=0.001$. Initial conditions are $S(0)=0.9$,
$I(0)=1-S(0)$ and $P(0)=10^{-6}$. (a) Long-time behavior of the
infected population (b) Boundary layer region showing the small-time
behavior of the infected population and pathogen number.}
\label{fig:smallRHO}
\end{figure*}
\noindent The quasi-steady state approximation employed above does
not imply that $\frac{dP}{dt}=0$, since $P$ will change in time
according to (\ref{approx}). A closed form solution is not possible
for the SIRS system. However, the long term behavior is
characterized by two possible equilibrium states. A disease-free
state $E_0=(1,0)$ and the endemic state $E_e$ where
\begin{displaymath}
E_e=(S_e,I_e)=\left(\frac{1}{R_0},\frac{(\phi+\psi)(R_0-1)}{R_0(\psi+1)}\right),
\end{displaymath}
and $P_e=I_e$ from (\ref{approx}), in agreement with (\ref{Pe}). It
is trivial to show by a linear analysis that, in the limit
$t\to\infty$, the SIRS system approaches this stable endemic state
when $R_0>1$, which is shown by a numerical solution in Figure
\ref{fig:smallRHO}(a). From (\ref{ics2}), the quasi-steady state
assumption requires $P(0)=I_0$ and the original initial conditions
(\ref{ics}) are not satisfied, except in the unique case where
$P_0=I_0$. This inconsistency arises because the approximation
(\ref{approx}) is only valid when $\frac{dP}{dt}\sim O(1)$ and
breaks down if $P$ changes rapidly at any point in the domain such
that $\rho\frac{dP}{dt}$ is no longer small. There is an initial
boundary layer for times of order $\rho$ in which the approximation
is invalid and the original scalings (\ref{scales}) are not
appropriate, Figure \ref{fig:smallRHO}(b). In this region
$\frac{dP}{dt}\gg1$ and we rescale the time in equations
(\ref{eqn1})-(\ref{Feqn}) using $t=\rho\tau$ to obtain equations for
this boundary layer region
\begin{align*}
\frac{dS}{d\tau}&=\rho\left[\phi(1-S)+\psi(1-S-I)-R_{0}PS\right],\\
\frac{dI}{d\tau}&=\rho\left[R_{0}PS-I\right],\\
\frac{dP}{d\tau}&=I-P.
\end{align*}
At leading order in $\rho$ we then have
\begin{displaymath}
\frac{dS}{d\tau}=0,\quad\quad
\frac{dI}{d\tau}=0,\quad\quad\frac{dP}{d\tau}=I-P,
\end{displaymath}
which has solution
\begin{displaymath}
S=S_0,\quad\quad I=I_0,\quad\quad P=I_0+(P_0-I_0)e^{-\tau}.
\end{displaymath}
The original initial condition $P(0)=P_0$ is now satisfied and as
$\tau\to\infty$ we exit the boundary layer and we have $P\to I_0$ in
agreement with the quasi-steady state approximation (\ref{ics2}). In
theory, unknown constants generated from the solution of the SIRS
model in the outer region would be fixed by matching the two
solutions in the limits as $t\to0$ and $\tau\to\infty$. In terms of
the long-term behaviour of an infectious disease, we can conclude
that the dynamics of the pathogen closely follow those of the
infected population and a standard SIRS model is sufficient for
a mathematical analysis. However, because the pathogen undergoes rapid changes, the initial transient interval may be important for short-term dynamics.\\

\subsection{Long-lived pathogenic state}
\label{Sec 2.2} If the pathogen survives significantly longer than
the duration of the infectious disease state, $T_P\gg T_I$, then
$\rho\gg1$. For convenience, in this section we set $\psi=0$ so that
immunity is permanent. This has the effect of slowing the rate at
which the susceptible compartment is repopulated. However, the
behavior of the solution is qualitatively similar and the simplified
system serves to highlight the primary features.
\begin{figure*}
\centering
\includegraphics[width=0.9\textwidth]{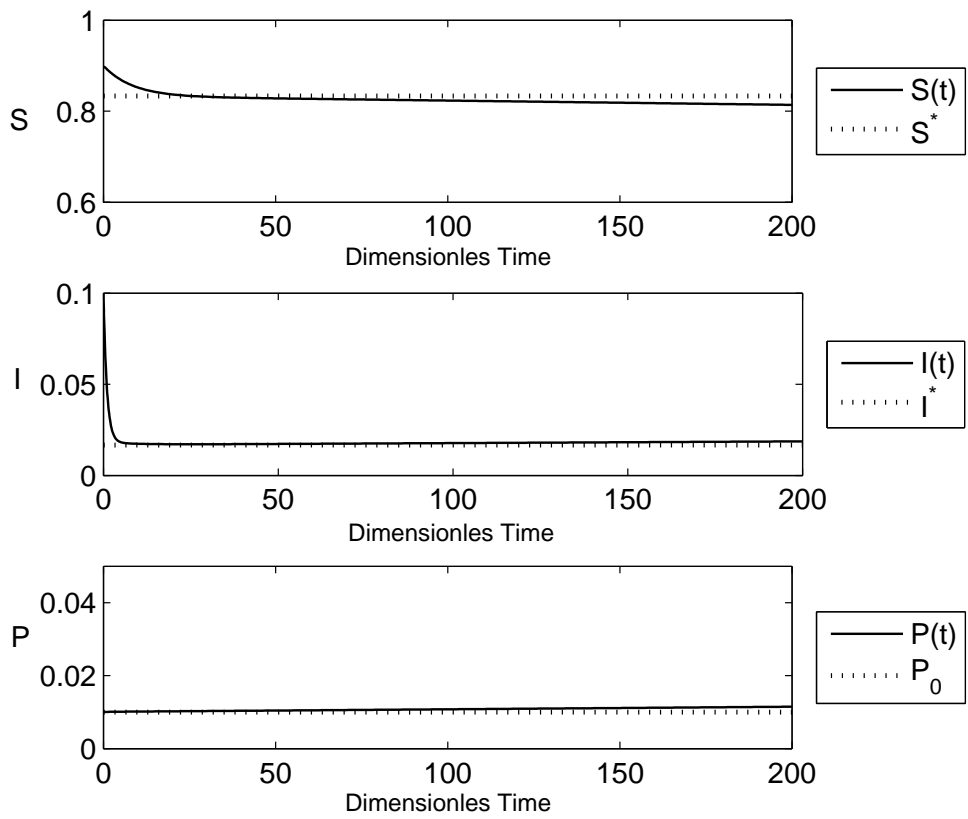}
\caption{Numerical solution  of the system
(\ref{eqn1})-(\ref{Feqn}), obtained with initial conditions
$S(0)=0.9$, $I(0)=0.1$ and $P(0)=0.01$. Parameter values are
$\phi=0.1$, $\psi=0$, $R_0=2$ and $\rho=1000$.} \label{fig:largeRHO}
\end{figure*}
The endemic equilibrium of the full system
(\ref{eqn1})-(\ref{Feqn}), with $\psi=0$, is now given by
\begin{equation}\label{Pe2}
\tilde{E}_e=(\tilde{S}_e,\tilde{I}_e,\tilde{P}_e)=\left(\frac{1}{R_0},\frac{\phi(R_0-1)}{R_0},\frac{\phi(R_0-1)}{R_0}\right),
\end{equation}
and the solution approaches $\tilde{E}_e$ as $t\to\infty$ when
$R_0>1$. When $\rho\gg1$ the right-hand side of equation
(\ref{Feqn}) is approximately zero, $(I-P)/\rho\approx0$, and we can
assume that $P$ is in a steady state such that $\frac{dP}{dt}=0$ and
$P=P_0$. The system reduces to
\begin{align*}
\frac{dS}{dt}&=\phi(1-S)-R_{0}P_0S,\\
\frac{dI}{dt}&=R_{0}P_0S-I,
\end{align*}
with initial conditions
\begin{displaymath}
S(0)=S_0,\quad\quad I(0)=I_0.
\end{displaymath}
The linear system is easily solved to obtain
\begin{align}
S(t)&=S^*+(S_0-S^*)e^{-\frac{\phi}{S^*}t},\label{sol1}\\
I(t)&=I^*+I^*\frac{S_0-S^*}{S^*-\phi}(e^{-\frac{\phi}{S^*}t}-e^{-t})+(I_0-I^*)e^{-t},\label{sol2}
\end{align}
where
\begin{equation}\label{Pstar}
S^*=\frac{\phi}{\phi+R_0P_0},\quad\quad I^*=R_0P_0S^*.
\end{equation}
\begin{figure*}
\centering
\includegraphics[width=0.9\textwidth]{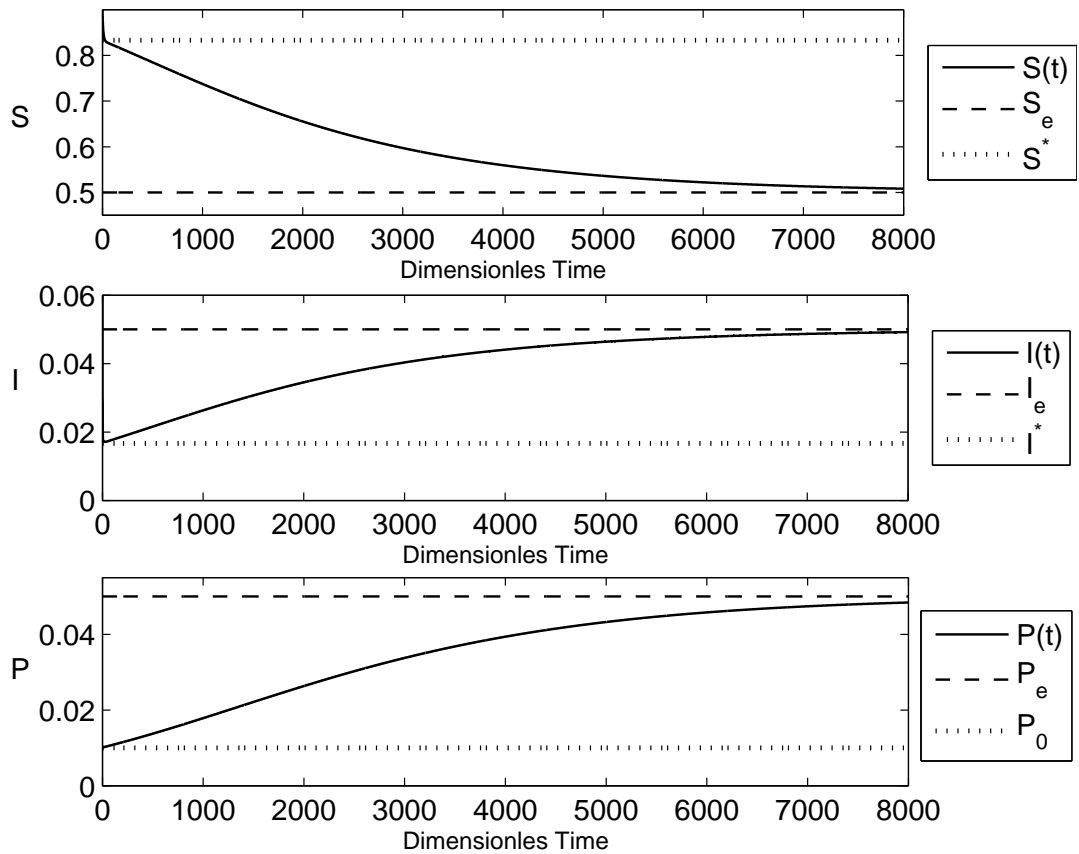}
\caption{Numerical solution  of the system
(\ref{eqn1})-(\ref{Feqn}), obtained with initial conditions
$S(0)=0.9$, $I(0)=0.1$ and $P(0)=0.01$. Parameter values are
$\phi=0.1$, $\psi=0$, $R_0=2$ and $\rho=1000$.}
\label{fig:largeRHO2}
\end{figure*}
The solution of the reduced system clearly satisfies the initial
conditions of the full system (\ref{ics}) as $t\to0$. However, as
$t\to\infty$ the solution does not approach the endemic equilibrium
(\ref{Pe2}). At relatively large times, the solution approaches an
alternative endemic state $E^*=(S^*,I^*,P_0)$, Figure
\ref{fig:largeRHO}. This inconsistency arises because, for times
$t\gtrsim O(\rho)$, $P$ is no longer at a steady state and begins to
grow towards its $\tilde{E}_e$ value, Figure \ref{fig:largeRHO2}.
Under such circumstances the steady state assumption breaks down. To
describe this large-time behaviour, we rescale the time
$t=\rho\theta$ and the system becomes
\begin{align*}
\frac{1}{\rho}\frac{dS}{d\theta}&=\phi(1-S)-R_{0}P_0S,\\
\frac{1}{\rho}\frac{dI}{d\theta}&=R_{0}P_0S-I,\\
\frac{dP}{d\theta}&=I-P.
\end{align*}
Now, for $\rho\gg1$ this can be approximated by
\begin{align*}
&\phi(1-S)-R_{0}PS=0,\\
&R_{0}PS-I=0,\\
&\frac{dP}{d\theta}=I-P,
\end{align*}
which has solution
\begin{equation}\label{implicitsol}
S(\theta)=\frac{\phi}{\phi+R_0 P(\theta)},\quad\quad
I(\theta)=\frac{\phi R_0 P(\theta)}{\phi+R_0 P(\theta)},
\end{equation}
where $P(\theta)$ satisfies the first-order equation
\begin{equation}\label{implicitF}
\frac{dP}{d\theta}=\frac{\phi R_0 P(\theta)}{\phi+R_0
P(\theta)}-P(\theta).
\end{equation}
For this large time solution to match the inner one,
(\ref{sol1})-(\ref{sol2}), we require
\begin{displaymath}
S(\theta\to 0)=S^*,\quad\quad I(\theta\to 0)=I^*,\quad\quad
P(\theta\to 0)=P_0.
\end{displaymath}
which yields
\begin{displaymath}
S(\theta=0)=\frac{\phi}{\phi+R_0 P_0},\quad\quad
I(\theta=0)=\frac{\phi R_0 P_0}{\phi+R_0 P_0},
\end{displaymath}
in agreement with (\ref{Pstar}). Equation (\ref{implicitF}) cannot
be solved explicitly for $P(\theta)$, however, an implicit solution
can be obtained
\begin{displaymath}
\frac{P_0}{P}\left(\frac{\phi+R_0P-\phi R_0}{\phi+R_0 P_0-\phi
R_0}\right)^{R_0}=e^{-\theta(R_0-1)}.
\end{displaymath}
Now, as $\theta\to\infty$ we find that the limiting value of
$P(\theta)$, which we denote as $P_\infty$, must satisfy
\begin{displaymath}
\left(\frac{\phi+R_0P_\infty-\phi R_0}{\phi+R_0 P_0-\phi
R_0}\right)^{R_0}=0,
\end{displaymath}
which yields
\begin{displaymath}
P_\infty=\frac{\phi(R_0-1)}{R_0}=\tilde{P}_e.
\end{displaymath}
It follows immediately from (\ref{implicitsol}) that $S\to
\tilde{S}_e$ and $I\to \tilde{I}_e$ as $\theta\to\infty$ and the
large time
solution approaches the endemic equilibrium (\ref{Pe2}) as required, Figure \ref{fig:largeRHO2}.\\

\subsection{Applications of the model}
\label{applications} The model can be used to describe the airborne
spread of a respiratory disease by aerosol droplets. The total
number of airborne pathogens is $P(t)=N_pD(t)$, where $D(t)$ is the
total number of droplets with active pathogens and $N_p$ is the
number of pathogens per droplet at the time of expulsion
(Stilianakis and Drossinos, 2010). The lifespan of droplets is identical to that of
pathogens ($T_P=T_D=1/\alpha$) as droplets and pathogens are
removed by identical processes since they are intrinsically linked.
Gravitational settling removes droplets from the environment, which
invariably results in the removal of the droplet's pathogen load.
Similarly, the inhalation of a droplet by a population member
removes pathogens, however, this process had little impact on
droplet numbers and can be neglected (Robinson et al. 2012). In addition, a
droplet is removed from the infection pathway through pathogen
inactivation (i.e. an airborne droplet carrying inactivated
pathogens cannot cause infection). The dimensional system describing
the infection dynamics is
\begin{align*}
\frac{dS}{dt}&=d(N-S)+\sigma(N-S-I)-\frac{\beta_d}{N}DS,\\
\frac{dI}{dt}&=\frac{\beta_d}{N}DS-(\mu+d)I,\\
\frac{dD}{dt}&=\kappa_d I-\alpha D,
\end{align*}
where $\beta_d=\beta N_p$ and $\kappa_d=\frac{\kappa}{N_p}$ are the
\textit{droplet} transmission and generation rates respectively.
Scaling $D\sim\frac{\kappa_d}{\alpha}N$, the dimensionless system is
\begin{align}
\frac{dS}{dt}&=\phi(1-S)+\psi(1-S-I)-R_0DS,\label{dropleteqn1}\\
\frac{dI}{dt}&=R_0DS-I,\label{dropleteqn2}\\
\rho\frac{dD}{dt}&=I-D,\label{dropleteqn3}
\end{align}
where all other scales and dimensionless numbers are defined as
before. For the particular case of influenza the infectious period
is approximately $5$ days, $\mu=0.2$/day, and an average human
lifespan is $70$ years, $d=4\times10^{-5}$/day. Airborne droplets
(of diameter $4\ \mu$m) are removed, through the processes of
gravitational settling and pathogen inactivation, at a rate of
$\alpha=37.44$/day (Robinson et al. 2012). The droplet generation rate $\kappa_d$ is based on the number of pathogen loaded droplets emitted during a cough. Generation rates per cough are taken as $160$/day (Nicas et al. 2005). The daily generation rate is then obtained by considering a 200-fold increase for a sneeze (Nicas et al. 2005) and a total of 11 sneezes and 360 coughs per day (Atkinson and Wein, 2008), which yields $\kappa_d=4.1\times10^{5}$/day. Loss of immunity is typically
associated with the emergence of new viral strains, with previously
infected individuals reverting to a susceptible state after
approximately $5$ years, $\sigma=5.5\times 10^{-4}$/day
(Truscott et al. 2012). \\

For an epidemic in progress the basic reproduction number is typically estimated from incidence data and can vary significantly between different pandemic and seasonal outbreaks. Estimates for the 2009 H1N1 pandemic are in the range $1.3-1.7$ (Yang et al. 2009) and an average seasonal value is approximately $1.3$ (Chowell et al. 2008). However, we have developed an explicit expression for the transmission rate per droplet $\beta_d$ which allows a direct estimation of the reproduction number. Firstly, it is important to note that the infectious agent is not the droplet but the pathogens it carries. Therefore, the transmission rate per droplet $\beta_d$ will depend the transmission rate per pathogen, $\beta_d=\beta_pq_dN_p$, where $q_d$ is the probability of deposition in the human respiratory tract. The minimum infectious dose required to transmit the infection is thus implicitly incorporated into the transmission rate. The transmission rate per pathogen $\beta_p$ is determined from the contact rate $c_d$ of a susceptible with a droplet and the probability $p_d$ that such a contact will result in successful transmission $\beta_p=c_dp_d$. To derive the contact rate with a droplet it is assumed that each infected person is surrounded by a droplet cloud with volume $V_{cl}$. It is further assumed that a susceptible individual comes in contact with a droplet through breathing during an encounter with this droplet cloud. If the average breathing rate is $B$ and $\tau_{ct}$ is a characteristic time of breathing during the encounter then the contact rate $c_d$ can be expressed as $c_d=c\frac{B}{V_{cl}}\tau_{ct}$, where $c$ is the average number of total contacts
a susceptible individual has per unit time. The transmission rate per droplet is thus
\begin{displaymath}
\beta_{d}=c\frac{B}{V_{cl}}\tau_{ct}p_dq_dN_p,
\end{displaymath}
and the number of pathogens per droplet can be determined by $N_p=V_d\rho_p$, where $V_d$ is the volume of the (spherical) pre-evaporative droplet and $\rho_p$ is the pathogen concentration of the lung fluid. All the relevant parameter values for influenza are discussed by Stilianakis and Drossinos (2010), and the corresponding values for a $4\mu$m droplet are summarized in Table \ref{table:Parameters}. The transmission rate per pathogen can now be calculated as $0.028$/day and the transmission rate per droplet then evaluates to $\beta_{d}=2.45\times10^{-5}$/day. Finally, the basic reproduction number can be estimated as $R_0\approx1.3$, in agreement with seasonal estimates from the literature. Using the parameter values discussed above the dimensionless numbers are estimated as
\begin{equation}\label{fluparameters}
R_0=1.3,\quad\quad \rho\approx0.005, \quad\quad \phi\approx0.0002,
\quad\quad \psi\approx0.003.
\end{equation}

\begin{table}[tp]
\begin{center}
\selectfont{\footnotesize{\begin{tabular}{ l l l l l} \hline
\multicolumn{2}{l}{parameter}  & \multicolumn{2}{l}{value} & \\ \hline
$c$ & contact rate & \multicolumn{2}{l}{13 per day}&   \\
$\rho_{p}$ & pathogen concentration in the lung fluid & \multicolumn{2}{l}{$3.71\times10^{6}$ pathogens cm$^{-3}$} &   \\
$B$ & breathing rate & \multicolumn{2}{l}{$24$ m$^{3}$ per day} &  \\
$V_{cl}$ & personal-cloud volume of an infected person & \multicolumn{2}{l}{$8$ m$^{3}$} &  \\
$p_{d}$ & infection probability by an inhaled pathogen & \multicolumn{2}{l}{0.052} &   \\
$\tau_{ct}$ & characteristic breathing (contact) time & \multicolumn{2}{l}{20 min} &   \\
$V_d$ & pre-evaporation (spherical) droplet volume & \multicolumn{2}{l}{$2.68\times 10^{-10}$ cm$^{3}$} &  \\
$q_{d}$ & inhaled droplet deposition probability & \multicolumn{2}{l}{$0.88$} &  \\ \hline
\end{tabular}}}
\end{center}
\caption{Parameter values used to determine the transmission rate per droplet.}
\label{table:Parameters}
\end{table}

\begin{figure*}
\includegraphics[width=1\textwidth]{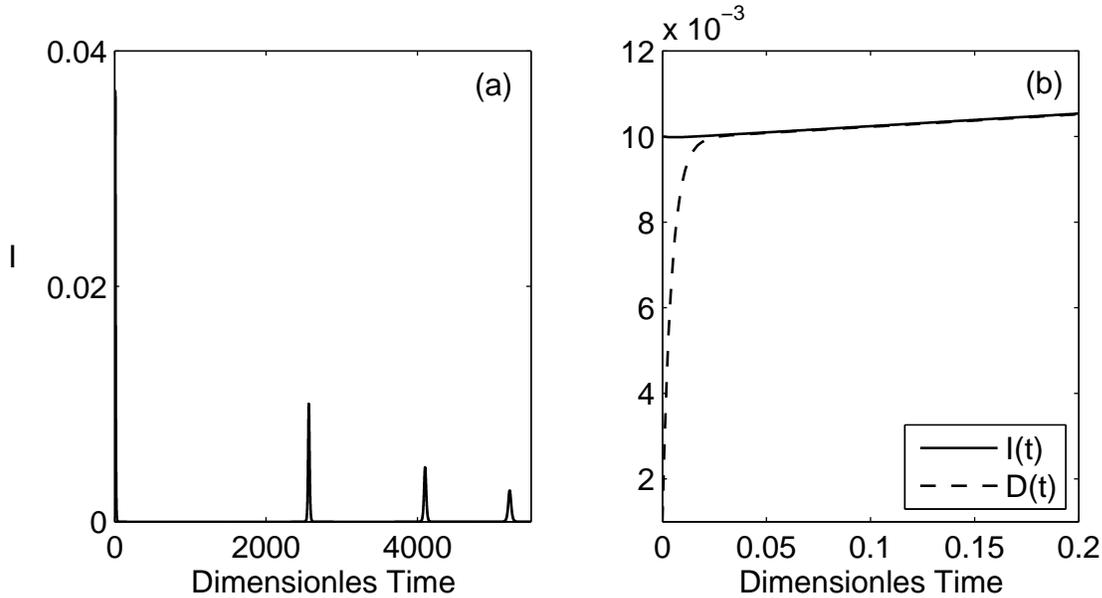}
\caption{Numerical solution of
(\ref{dropleteqn1})-(\ref{dropleteqn3}), with initial conditions
$S(0)=0.99$, $I(0)=0.01$ and $D(0)=10^{-3}$. Parameter values for
influenza are $R_0=1.3$, $\rho\approx0.005$, $\phi\approx0.0002$ and
$\psi\approx0.003$.} \label{fig:influenza}
\end{figure*}

The lifespan of droplets (and pathogens) is much less than that of
an infected person such that $\rho\ll1$. Also, for influenza,
$\phi\ll\psi\ll1$ and the repopulation of the susceptible class
following an outbreak is a slow process, primarily driven by loss of
immunity and not natural births. The numerical solution approaches
the endemic equilibrium with damped oscillations of large period
$\sim10$ years, Figure \ref{fig:influenza}(a). The magnitude of the
period is highly sensitive to the duration of immunity, with much
shorter times observed between outbreaks if $1/\sigma$ is
reduced. The small-time behavior of the model, where droplets
rapidly achieve a balance with the infected population, is shown in
Figure \ref{fig:influenza}(b). After approximately $6$ hours
droplets and
infected individuals achieve a balance and droplet dynamics thereafter follow the infected population.\\

\begin{figure*}
\includegraphics[width=1\textwidth]{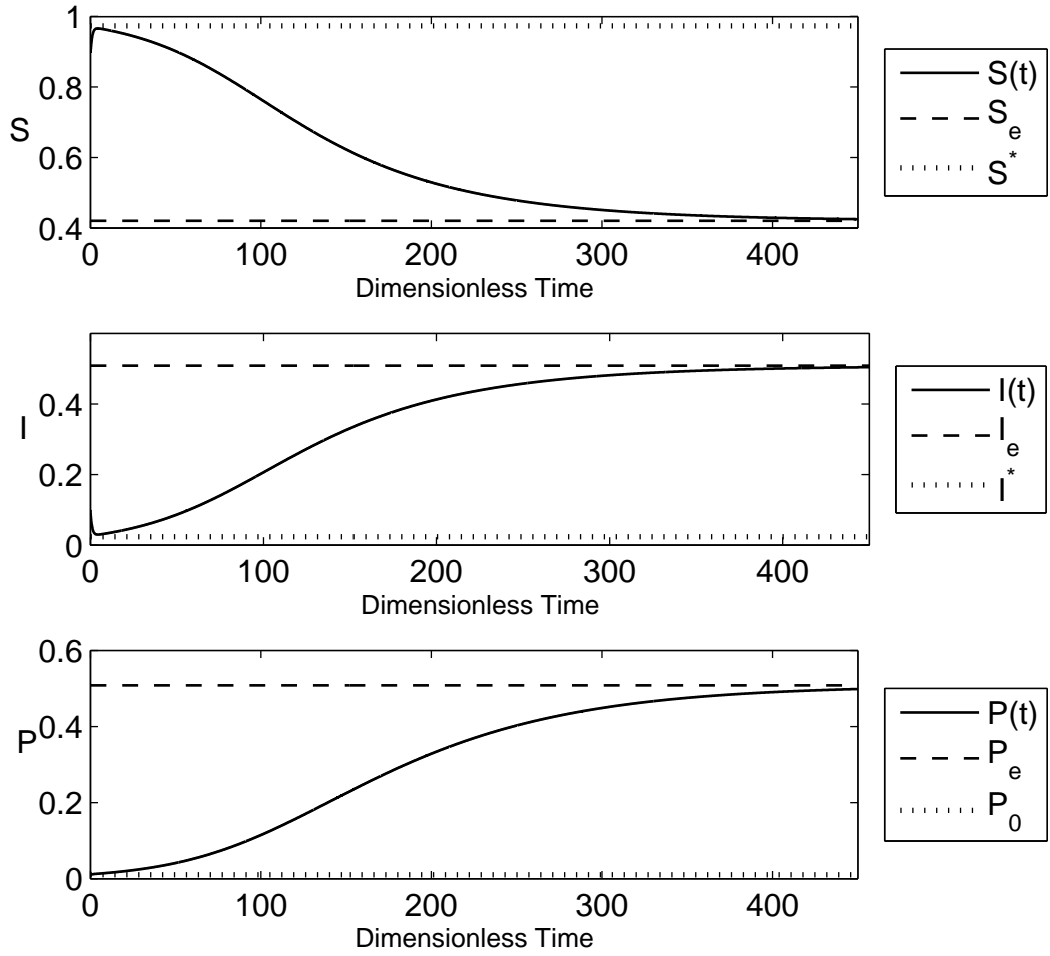}
\caption{Numerical solution of (\ref{eqn1})-(\ref{Feqn}), with
initial condition $S(0)=0.9$, $I(0)=0.1$ and $P(0)=0.01$. Parameter
values for cryptosporidium are $R_0=2.38$, $\rho\approx46.7$,
$\phi\approx2.7\times 10^{-4}$ and $\psi\approx7.1$.}
\label{fig:cryptosporidium}
\end{figure*}

The model can also be used to describe the waterborne route of
transmission (Code\c{c}o, 2001, Tien and Earn, 2010). An example of a pathogen that
causes diarrhoeal disease in humans is the protozoa Cryptosporidium.
Large outbreaks of Cryptosporidium due to contaminated drinking
water are common (Eisenberg et al. 1998, Glaberman et al. 2002). The variable $P(t)$
will represent the number of oocysts in a water source. This
pathogen has been recorded surviving in an aqueous suspension for up
to 12 months in cold temperatures $\alpha=0.003$/day,
(Peeters et al. 1989). An infection typically persists for $6-9$ days,
$\mu=0.14$/day,  with infected individuals shedding up to $10^{5-7}$
oocysts per gram of faeces (Medema et al. 2009). With an average human faecal
production rate of 106 g/day (Cummings et al. 1992), we set
$\kappa=10^7$/day. The magnitude of the transmission rate by the
waterborne route (in the absence of direct person-to-person
transmission) has been estimated to be in the range
$[10^{-11},10^{-7}]$ and we take $\beta=10^{-10}$/day
(Eisenberg et al. 1998). No immunity is acquired following infection and,
for convenience, we assume that immunity is lost one day after
recovery $\sigma=1$/day. Using these parameter values we estimate
dimensionless numbers as
\begin{displaymath}
R_0=2.38,\quad\quad \rho=46.7,\quad\quad \phi=2.7\times
10^{-4},\quad\quad \psi=7.1.
\end{displaymath}
The long lifespan of the cryptosporidium oocysts yields $\rho\gg1$.
The solution of the model is plotted in Figure
\ref{fig:cryptosporidium}. An initial rapid period can be observed
during which $P$ is approximately constant and $S$ and $I$ rapidly
approach an intermediate state $(S^*,I^*)=(0.97,0.02)$. The observed
growth in the susceptible population in this region is an artifact
of the choice of initial condition. At large times the solution
converges to the endemic state $E_e=(S_e,I_e,P_e)=(0.42,0.5,0.5)$.

\section{Seasonal variation in pathogen inactivation}
\label{Sec3} In this section we consider the dynamics of the model
presented in Section \ref{Sec2} when seasonal forcing is applied to
the pathogen inactivation rate. We make no assumption as to the
precise source of the forcing, merely that it manifests through
increased pathogen survival in winter.
We consider the specific case of the airborne transmission of respiratory infections by aerosol droplets, as described in Section \ref{applications}.\\

\subsection{The quasi-steady state}
\label{sec 3.1} To describe the seasonal cycle of respiratory
infectious diseases we consider the effects of applying a seasonal
forcing to the droplet model of Section \ref{applications}. The
forcing is implemented via the droplet removal rate. The dimensional
model is
\begin{align*}
\frac{dS}{dt}&=d(N-S)+\sigma(N-S-I)-\frac{\beta_d}{N}DS,\\
\frac{dI}{dt}&=\frac{\beta_d}{N}DS-(\mu+d)I,\\
\frac{dD}{dt}&=\kappa_d I-\alpha(t)D.
\end{align*}
We take
\begin{displaymath}
\alpha=\theta+\eta_{0}\{1+\eta_{1}\cos(\omega t)\},
\end{displaymath}
where $\omega$ is the frequency of oscillation of the inactivation
rate, $\theta$ is the constant rate at which gravity and inhalation
remove droplets. The parameter $\eta_0$ is the pathogen inactivation
rate in the absence of seasonal forcing and $\eta_1$ represents the
amplitude of the applied forcing term. For convenience, we denote
$\alpha_0=\theta+\eta_0$. We nondimensionalise by scaling
\begin{displaymath}
S,I\sim N,\quad
D\sim\frac{\kappa_d}{\alpha_0}N,\quad\alpha\sim\alpha_0,\quad
t\sim\frac{1}{\mu+d},
\end{displaymath}
and we can write the dimensionless system as
\begin{align}
\frac{dS}{dt}&=\phi(1-S)+\psi(1-S-I)-R_{0}DS,\label{seasonaleqn1}\\
\frac{dI}{dt}&=R_{0}DS-I,\label{seasonaleqn2}\\
\rho\frac{dD}{dt}&=I-\alpha D,\label{seasonaleqn3}\\
\alpha&=1+\alpha_{1}\cos(\Omega t),\label{seasonaleqn4}
\end{align}
where dimensionless numbers are defined as
\begin{equation}\label{seasonal_dimen_params}
R_{0}=\frac{\beta_d\kappa_d}{\alpha_0(\mu+d)},\quad\phi=\frac{d}{\mu+d},\quad\psi=\frac{\sigma}{\mu+d},\quad\rho=\frac{\mu+d}{\alpha_0}=\frac{T_D}{T_I},\quad\alpha_1=\frac{\eta_0\eta_1}{\alpha_0},\quad\Omega=\frac{\omega}{\mu+d}.
\end{equation}
In the absence of seasonal forcing ($\alpha_1\equiv0$) the long-term
behaviour is characterized by an endemic equilibrium state given by
\begin{displaymath}
S_e=\frac{1}{R_0},\quad\quad
I_e=D_e=\frac{(\phi+\psi)(R_0-1)}{R_0(\psi+1)},
\end{displaymath}
in agreement with (\ref{Pe}) where $\alpha=\alpha_0$. This equilibrium exists and is stable when $R_0>1$ and one of either $\phi$ or $\psi$ are nonzero, enabling the repopulation of the susceptible compartment.\\

The lifespan of airborne respiratory pathogens is relatively small,
as gravity and pathogen inactivation invariably remove the
pathogen-loaded droplets on a timescale significantly shorter than
the infectious period, $T_D\ll T_I$, and we can reasonably assume
$\rho\ll1$. We adopt the quasi-steady state assumption and
approximate the number of droplets by
\begin{displaymath}
D=\frac{I}{\alpha}.
\end{displaymath}
In addition, unless immunity is permanently acquired following
infection, the duration of the immune period is small compared with
the average human lifespan and $\phi\ll\psi$. We thus allow
repopulation of the susceptible class solely through the loss of
immunity and neglect the natural birth process, i.e. $\phi\approx0$.
The model reduces to two equations
\begin{align}
\frac{dS}{dt}&=\psi(1-S-I)-\frac{R_{0}}{\alpha}IS,\label{Seqn}\\
\frac{dI}{dt}&=\frac{R_0}{\alpha}IS-I,\label{Deqn}
\end{align}
with $\alpha=1+\alpha_{1}\cos(\Omega t)$. In contrast to models with
sinusoidal forcing applied directly to the transmission rate
$\beta$, we find that the forcing appears in the denominator of the
nonlinear term. Singularities are possible if $\alpha=0$, however, a
small forcing amplitude $\eta_1\ll1$ always yields $\alpha_1\ll1$
and division by zero is avoided.

\subsection{Small-amplitude seasonal forcing}
\label{sec 3.2} Equation (\ref{Deqn}) allows $S$ to be written in
terms of $I$ as
\begin{displaymath}
S=\frac{\alpha(\dot{I}+I)}{R_{0}I},
\end{displaymath}
where the dot denotes the time derivative. This can be substituted
into (\ref{Seqn}) to yield a single equation for $I$,
\begin{align*}
\alpha I\ddot{I}+\left[\dot{\alpha}I-\alpha\dot{I}+\alpha\psi
I+R_{0}I^2\right]\dot{I}+I^2\left[\dot{\alpha}-\psi R_0+\psi R_0
I+R_0 I +\alpha\psi\right]=0.
\end{align*}
Assuming the seasonal forcing is small ($\alpha_1\ll1$), a
perturbation $\xi(t)$ to the endemic equilibrium $I_e$ such that
\begin{displaymath}
I\approx I_e+\alpha_1\xi(t)=I_s(t),
\end{displaymath}
satisfies the second order linear inhomogeneous equation
\begin{equation}\label{xidiffeqn}
\frac{d^2\xi}{dt^2}+(\psi+R_0I_e)\frac{d\xi}{dt}+\psi(R_0-1)\xi=I_e\left[\Omega\sin{(\Omega
t)}-\psi\cos{(\Omega t)}\right].
\end{equation}
The characteristic equation of the corresponding homogeneous problem
is
\begin{displaymath}
\lambda^{2}+(\psi+R_0I_e)\lambda+\psi(R_0-1)=0,
\end{displaymath}
which has two negative roots $\lambda_{1,2}<0$ given by
\begin{displaymath}
\lambda_{1,2}=\frac{-(\psi+I_eR_0)\pm\sqrt{(\psi-I_eR_0)^2-4R_0I_e}}{2}.
\end{displaymath}
The solution of the inhomogeneous equation is then
\begin{equation}\label{xieqn}
\xi(t)=C_1e^{-|\lambda_1|t}+C_2e^{-|\lambda_2|t}+I_e\frac{\Omega[I_eR_0-\Omega^2-\psi^2]\sin{(\Omega
t)}-I_eR_0[\psi(\psi+1)+\Omega^2]\cos{(\Omega
t)}}{I_e^2R_0^2[(\psi+1)^2+\Omega^2]-2R_0\Omega^2I_e+\Omega^4+\psi^2\Omega^2},
\end{equation}
where $C_1$ and $C_2$ are arbitrary constants. At large times,
$\xi(t)$ approaches a periodic solution with period $\Omega$ and
amplitude
\begin{displaymath}
A=I_e\sqrt{\frac{\psi^2+\Omega^2}{I_e^2R_0^2[(\psi+1)^2+\Omega^2]-2R_0\Omega^2I_e+\Omega^4+\psi^2\Omega^2}}.
\end{displaymath}
In addition, the solution for $\xi$ has a phase $\Phi$ which will
satisfy
\begin{equation}\label{shift}
\tan{\Phi}=\frac{I_eR_0(\psi^2+\psi+\Omega^2)}{\Omega(\psi^2+\Omega^2-I_eR_0)}.
\end{equation}
Thus, for small-amplitude seasonal forcing, the infected population
can be approximated by
\begin{equation}\label{approxsol}
I_s(t)=I_e+\alpha_1 A\sin{(\Omega t+\Phi)}.
\end{equation}
Linear theory indicates that the infected population oscillates
about the endemic state $I_e$ with an amplitude $\alpha_1 A$. The
period of oscillation is identical to that of the applied seasonal
forcing and independent of the duration of acquired immunity.
However, the disease incidence and seasonally forced pathogen
inactivation are out of phase with each other. This is readily seen
by rewriting (\ref{approxsol}) in the form
\begin{equation}\label{approxsol2}
I_s(t)=I_e+\alpha_1
A\cos{\left\{\Omega\left(t+\frac{2\Phi-\pi}{2\Omega}\right)\right\}},
\end{equation}
and there is a time shift of $(2\Phi-\pi)/2\Omega$ between the two
quantities, whereby $I_s(t)$ achieves its peak value before $\alpha$
does when $2\Phi-\pi>0$. Furthermore, the maximum of $I_s(t)$ will not occur before the minimum of $\alpha(t)$ when
\begin{equation}\label{shiftcondition}
0\leq\frac{2\Phi-\pi}{2\Omega}\leq\frac{\pi}{\Omega}.
\end{equation}
Of course, a peak in $\alpha$ corresponds to the point of maximum pathogen inactivation
and thus to a minimum in pathogen survival, and logically peak
disease incidence should not coincide with minimal survival.
Therefore, a more interesting and informative quantity is the delay
between minimum $\alpha$ (peak pathogen survival) and peak disease
incidence. We denote this quantity as $\delta$ and find
\begin{equation}\label{delta}
\delta=\frac{3\pi-2\Phi}{2\Omega},
\end{equation}
where $\Phi$ can be determined from (\ref{shift}). Thus, the extent of the observed delay between peak survival and peak incidence is determined from the specific disease parameters and the period of the seasonally forced inactivation rate.\\

Another consideration in the relative positions of the maximum incidence and inactivation is the time of pathogen introduction, corresponding to the time the first infected individual is introduced into the population. For example, pathogen introduction at the point of peak survival could have radically different dynamics to an introduction at the time of peak inactivation. To address the consequences of the introduction time we consider a seasonal variation of the form
\begin{equation}\label{eq:InitialCondition}
\alpha (t) = 1 + \alpha_1 \cos \big [\Omega( t - t_0) \big ],
\end{equation}
and then investigate the dependence of our results on $t_0$. Equation~(\ref{eq:InitialCondition}) implies that the infected initial condition $I(t=0)$ may be introduced at different points during the seasonal variation of the pathogen inactivation rate. Namely, the infected person(s) at $t=0$ may be introduced in times that correspond to different pathogen inactivation rates. Given equation (\ref{eq:InitialCondition}), the calculations were repeated, and we found that the time shift $t - t_0$ appears in the inhomogeneous
term of the differential equation (\ref{xidiffeqn}) and the infected population then satisfies
\begin{displaymath}
I_s (t) = I_e + \alpha_1 A \sin \big [ \Omega (t - t_0) + \Phi \big ],
\end{displaymath}
which can be written as
\begin{equation}\label{approxsol3}
I_s (t) = I_e + \alpha_1 A \cos \Big \{ \Omega \big (t - t_0 + \frac{2 \Phi - \pi}{2 \Omega} \big ) \Big \}.
\end{equation}
It can then be easily shown that the delay $\delta$ does not depend on $t_0$, i.e. equation (\ref{delta}) remains valid, even for $t_0 \neq 0$. Therefore, we conclude that, while the solution of the differential equation will invariably depend on the time of pathogen introduction, the delay between incidence $I(t)$ and inactivation $\alpha(t)$ does not.

\subsection{The case of influenza}
\label{sec 3.3} As discussed in Section \ref{applications}, in the case of influenza we find $\phi\ll1$ and it is reasonable to neglect the natural birth process. The values of the dimensionless parameters $R_0$, $\rho$ and $\psi$ are given by equation (\ref{fluparameters}). The dimensionless frequency and amplitude are defined in (\ref{seasonal_dimen_params}). Assuming a seasonal forcing with a period of one year yields a frequency of $\Omega=0.086$. Droplet removal through inhalation is negligible and $\theta$ can be approximated by the gravitational settling rate alone, for a $4\mu$m droplet this yields $\theta=28.8$/day (Robinson et al. 2012). The pathogen inactivation rate is assumed to be independent of droplet size and is taken to be $\eta_0=8.64$/day (Hemmes et al. 1960). Finally, assuming a value of $\eta_1=0.1$ for the forcing amplitude yields $\alpha_1=0.023$. In reality, the value of $\eta_1$ depends on the seasonal driver (e.g. temperature, humidity) and its related seasonal cycle. This work merely assumes that $\eta_1$ is small and the chosen value determines the amplitude of the forcing term and thus the amplitude of the disease incidence. It is clear from (\ref{approxsol}) that the delay and period of the linearised solution are not affected by the choice of $\eta_1$.\\

\begin{figure*}[tp]
\centering
\includegraphics[width=0.6\textwidth]{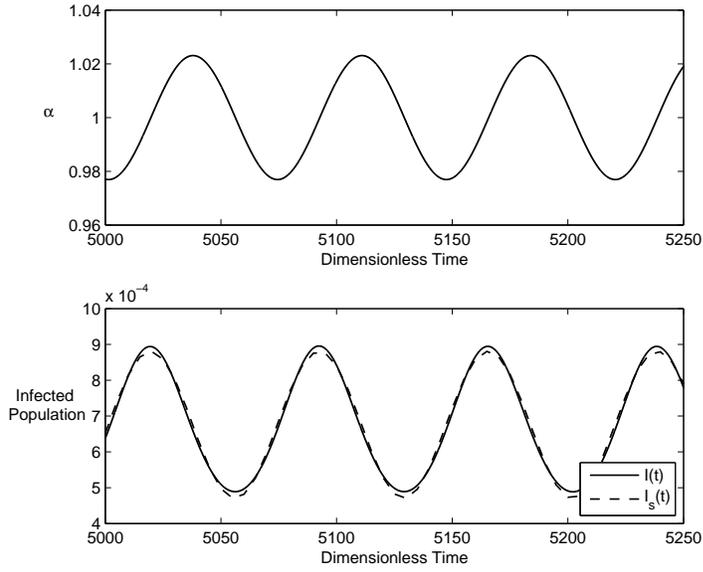}
\caption{Influenza seasonality with initial conditions $S(0)=0.99$,
$I(0)=0.01$ and $D(0)=10^{-3}$. Parameter values are $R_0=1.3$,
$\rho\approx0.005$, $\phi\approx0.0002$, $\psi\approx0.003$,
$\Omega=0.086$ and $\alpha_1=0.023$, with $t_0=0$.}
\label{fig:fluseasonality}
\end{figure*}
\begin{figure*}[htp]
\centering
\includegraphics[width=0.6\textwidth]{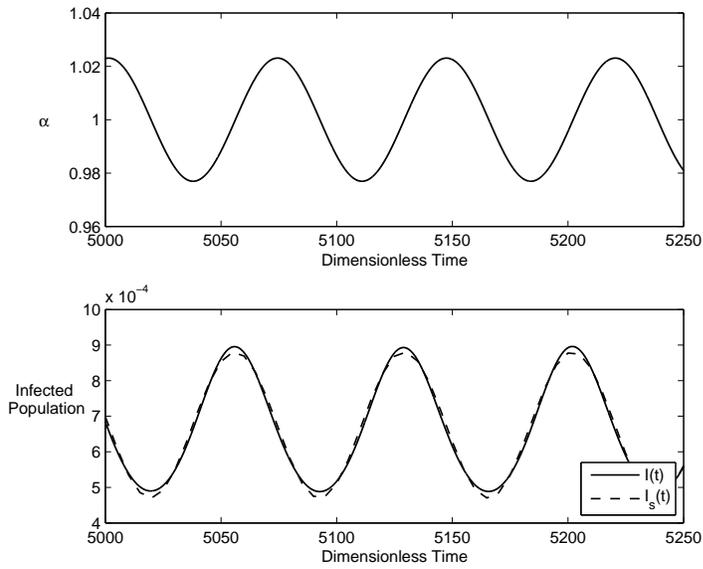}
\caption{Influenza seasonality with initial conditions $S(0)=0.99$,
$I(0)=0.01$ and $D(0)=10^{-3}$. Parameter values are $R_0=1.3$,
$\rho\approx0.005$, $\phi\approx0.0002$, $\psi\approx0.003$,
$\Omega=0.086$ and $\alpha_1=0.023$, with $t_0=\pi/\Omega$.} \label{fig:fluseasonality2}
\end{figure*}
The disease incidence $I_s(t)$ will reach its maximum before the maximal viral inactivation $\alpha(t)$ provided $2\Phi-\pi>0$. From (\ref{shift}), we calculate $\tan\Phi=0.0167$, the inverse of which yields $\Phi_f=0.0167$, where the subscript $f$ indicates that this angle lies in the fundamental interval $[-\pi/2,\pi/2]$. However, the trigonometric coefficients in (\ref{xieqn}) are both negative and $\Phi$ must lie in the third quadrant, yielding $\Phi=0.0167+\pi$. It follows that $2\Phi-\pi=3.175>0$. Furthermore, the relative positions of the incidence and inactivation curves can be further restricted, by requiring the satisfaction of (\ref{shiftcondition}), to ensure that the maximum of $I_s(t)$ will not occur before the minimum of $\alpha(t)$, and this condition reduces to
\begin{displaymath}
\frac{\pi}{2}\leq\Phi\leq\frac{3\pi}{2} \quad \Longrightarrow \quad - \frac{\pi}{2} \leq \Phi_f \leq \frac{\pi}{2},
\end{displaymath}
which is always satisfied by $\Phi_f$.\\

A numerical solution of the dimensionless system (\ref{seasonaleqn1})-(\ref{seasonaleqn4}), showing the
pathogen inactivation $\alpha(t)$, the disease incidence $I(t)$ and the solution of the linearised system $I_s(t)$, is shown in Figure
\ref{fig:fluseasonality}. The maximum disease incidence clearly occurs \textit{before} maximum inactivation and \textit{after} minimum inactivation.  The time delay between peak virus survival
(minimal $\alpha$) and peak disease incidence is clearly visible and
can be calculated from (\ref{delta}) as $\delta\approx18.07$. Both
curves, $I(t)$ and $\alpha(t)$, are observed to oscillate with the same period, as
predicted by the linear analysis. The linearized quasi-steady solution
$I_s(t)$, given by (\ref{approxsol3}), provides a very good
approximation to the solution of the full system. Figure 7 displays the disease incidence for the case $t_0=0$, such that the initial infected individual was introduced into the population at the point of maximum viral inactivation (minimum survival). To demonstrate the independence of the delay $\delta$ on the time of viral introduction, the simulation was re-run with the infected individual introduced at the point of minimum inactivation (maximum survival), i.e. $t_0=\pi/\Omega$, Figure 8. The timing of peak disease incidence changes, however the predicted delay remains constant.\\

\begin{figure*}[tp]
\centering
\includegraphics[width=0.8\textwidth]{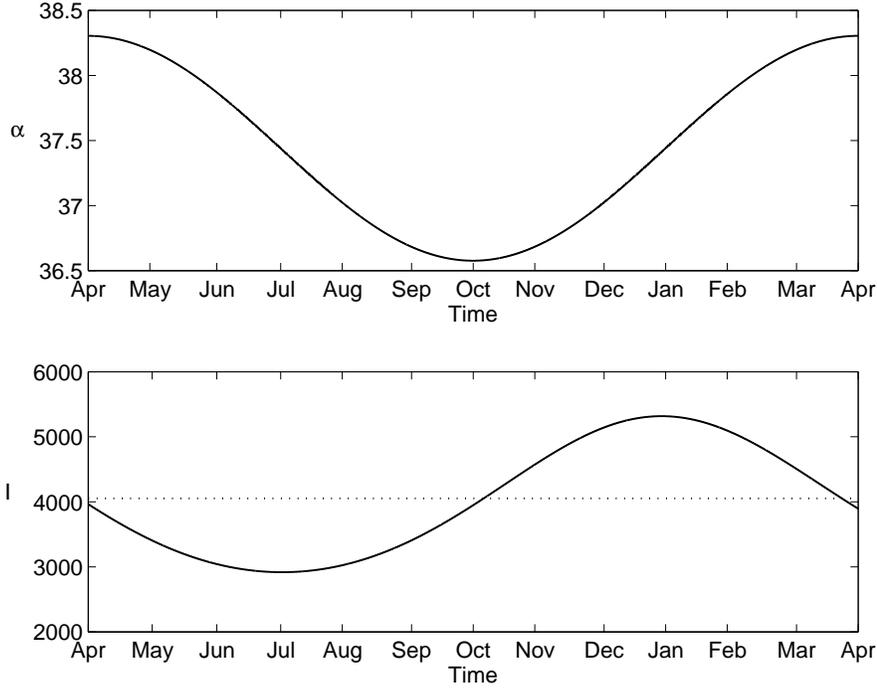}
\caption{Numerical solution obtained with initial condition
$S(0)=0.99$, $I(0)=0.01$ and $D(0)=10^{-3}$. Parameter values
$R_0=1.3$, $\rho\approx0.005$, $\phi\approx0.0002$,
$\psi\approx0.003$, $\Omega=0.086$ and $\alpha_1=0.023$. All
variables are dimensional.} \label{fig:dimensionalplot}
\end{figure*}

\begin{figure*}[tp]
\includegraphics[width=1\textwidth]{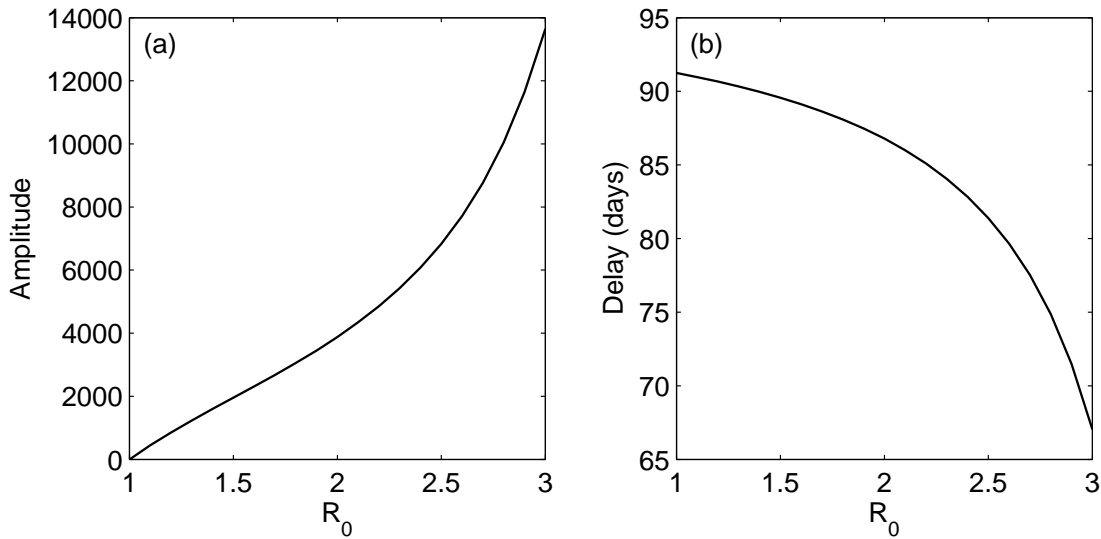}
\caption{(a) Dimensional amplitude ($=N\alpha_1A$) as a function of
$R_0$ (b) Dimensional delay ($=\delta/(\mu+d)$) as a function of
$R_0$. Parameter values $N=6\times10^6$, $d=4\times10^{-5}$/day,
$\mu=0.2$/day, $\theta=28.8$/day $\eta_0=8.64$/day, $\eta_1=0.1$,
$\omega=\frac{2\pi}{365}$.} \label{fig:VR0}
\end{figure*}

A numerical solution is shown in terms of dimensional variables in
Figure \ref{fig:dimensionalplot}, where it is assumed that the peak
in disease incidence occurs in January and, for illustration
purposes, we have fixed the total human population at $6$ million.
The solution indicates that disease incidence peaks approximately
three months ($\approx90.33$ days calculated from (\ref{delta}))
after the influenza virus inactivation reaches its minimum, at which
time virus survival is at its peak. The magnitude of this delay is
dependent on the disease parameters. The delay $\delta$ is plotted
in dimensional form for varying $R_0$ in Figure \ref{fig:VR0}(b). As
$R_0$ increases the delay decreases and the disease incidence peak
occurs earlier after the peak in influenza virus survival. The
amplitude of the infectious wave as a function of $R_0$ is plotted
in Figure \ref{fig:VR0}(a). Clearly, amplitude increases with
increasing $R_0$, indicating the occurrence of more severe outbreaks
for large $R_0$ values.




\section{Discussion}
\label{Sec4}
In this work we have presented a model for the transmission of an infectious
disease by means of a pathogenic state capable of surviving in an environmental
reservoir outside of its host organism. For a reproduction number exceeding unity,
$R_0>1$, the pathogen dynamics are found to depend on the magnitude of a dimensionless
number $\rho$, which represents the ratio between the pathogen lifespan and that
of an infected individual. If the duration of the infectious disease state is
significantly longer than the lifespan of the pathogen (e.g. for influenza), yielding
$\rho\ll1$, a rapid transient period is observed during which pathogen and infected
numbers rapidly achieve a balance. Thereafter, the pathogen dynamics identically follow
those of the infected population. At large times, the solution approaches a stable endemic
disease state, exhibiting damped oscillations as it does so. In contrast, if the pathogen
is capable of surviving significantly longer than the duration of the infectious disease
state (e.g. cryptosporidium), yielding $\rho\gg1$, the susceptible and infected populations
first approach an intermediate state during which the pathogen number remains approximately
constant. However, at large times all variables converge to a stable endemic state.\\

The effect of incorporating seasonal forcing through the pathogen inactivation rate was
considered for the specific case of a respiratory infection mediated by airborne
pathogen-loaded droplets. A quasi-steady state approximation allows the droplet dynamics
to be neglected resulting in a classic SIRS model with the seasonal forcing appearing in
the denominator of the transmission term. The assumption of small-amplitude seasonal forcing
yields a linearized approximation to the infected population, given by (\ref{approxsol2}).
It was found that the infected population oscillates about the endemic disease state with an
identical period to that of the seasonally forced pathogen inactivation rate and the period
is independent of the duration of acquired immunity. In addition, the peak in pathogen survival
(corresponding to minimal pathogen inactivation) was found to precede peak disease incidence.
For the particular case of influenza with a reproduction number of $R_0=1.3$, this phase shift
was calculated to be approximately three months, with shorter delays observed for larger $R_0$ values, Figure \ref{fig:VR0}.\\

A final topic worth noting in relation to the seasonal model presented in Section \ref{Sec3} is the interpretation of the basic reproduction number. In the seasonal context, this no longer corresponds to the average number of secondary infections per primary infection since this value will invariably be dependent on time. Attempts have been made to define $R_0$ in this context (Baca\"{e}r and Ait Dads 2012, Grassley and Fraser 2006) and the various attempts to derive a time-dependent expression have focused on the analysis of a time-dependent transmission term. However, in this work we attempt to tackle the issue of seasonality from a novel perspective by linking the seasonal signal to a palpable physical process. As such, the time-dependent $R_0$ would be intrinsically linked to the pathogen inactivation rate which is a complex function of environmental factors, which in turn are highly variable with geographic location. To this end, an elaboration on the issue of time-dependence is beyond the scope of this work.\\

The exact mechanism driving the seasonal occurrence of infectious
diseases is uncertain. Several factors undoubtedly impact the
dynamics, including human behavior and pathogen survival, and many
of these factors are difficult to quantify. In particular, the
burden placed on health systems by seasonal influenza, and the
constant threat of deadly pandemics, makes understanding the
transmission mechanisms of the virus an important issue.
Experimental studies have identified the impact of environmental
conditions on virus survival as a crucial factor. However, the
primary sources of viral inactivation (temperature, humidity, solar
radiation, etc.) have been difficult to identify, and indeed a
complex combination of these sources working in unison is a likely
culprit. More rigorous experimental work is needed to ascertain the
primary source of the observed seasonality. Our model provides an
insight into how annual variations in virus inactivation can
influence disease incidence and the model is sufficiently general
that it can be applied to various transmission pathways. A better
understanding of pathogen removal processes could provide the
opportunity to develop more complex models incorporating the effect
and interaction of multiple environmental factors.



\begin{thebibliography}{7}
\bibitem{anderson2} Anderson, R.M., May, R.M., 1981. The population dynamics of microparasites and their invertebrate hosts. Phil. Trans. Roy. Soc. London B. 291, 451-524.

\bibitem{anderson} Anderson, R.M., May, R.M., 1991. Infectious diseases in humans: dynamics and control. Oxford University Press.

\bibitem{atkinson} Atkinson, M.P., Wein, L.M., 2008. Quantifying the routes of transmission for pandemic influenza. Bull. Math. Biol. 70, 820-867.

\bibitem{bacaer} Baca\"{e}r, N., Ait Dads, E.H., 2012. On the biological interpretation of a definition for the parameter $R_0$ in periodic population models. J. Math. Biol. 65, 601-621.

\bibitem{bertuzzo} Bertuzzo, E., Casagrandi, R., Gatto, M., Rodriguez-Iturbe, I., Rinaldo, A., 2010. On spatially explicit models of cholera epidemics, J. R. Soc. Interface. 7, 321333.

\bibitem{burie} Burie, J.B., Calonnec, A., Ducrot, A., 2006. Singular perturbation analysis of travelling waves for a model in phytopathology. Math. Model. Nat. Phenom. 1, 49-62.

\bibitem{cannell} Cannell, J.J., Zasloff, M., Garland, C.F., Scragg, R., Giovannucci, E., 2008. On the epidemiology of influenza. Virology Journal. 5, 29.

\bibitem{chowell} Chowell, G., Miller, M.A., Viboud, C., 2008. Seasonal influenza in the United States, France and Australia: transmission and prospects for control. Epidemiol. Infect. 136, 852-864.

\bibitem{codeco} Code\c{c}o, C.T., 2001. Endemic and epidemic dynamics of cholera: the role of the aquatic reservoir. BMC Infect. Dis. 1, 1.

\bibitem{cummings} Cummings, J.H., Bingham, S.A., Heaton, K.W., Eastwood, M.A., 1992. Fecal weight, colon cancer risk, and dietary intake of nonstarch polysaccharides (dietary fiber). Gastroenterology. 103, 1783-1789.

\bibitem{dowell01} Dowell, S.F., 2001. Seasonal variation in host susceptibility and cycles of certain infectious diseases. Emerg. Infect. Dis. 7, 369-374.

\bibitem{dushoff} Dushoff, J., Plotkin, J.B., Levin, S.A., Earn, D.J.D., 2004. Dynamical resonance can account for seasonality of influenza epidemics. PNAS. 101, 16915-16916.

    \bibitem{eisenberg} Eisenberg, J.N., Seto, E.Y., Colford Jr, J.M., Olivieri, A., Spear, R.C., 1998. An analysis of the Milwaukee Cryptosporidiosis outbreak based on a dynamic model of the infection process. Epidemiology. 9, 255-263.

\bibitem{fisman} Fisman, D.N., 2007. Seasonality of infectious diseases. Annu. Rev. Public Health. 28, 127-143.

\bibitem{glaberman} Glaberman, S., Moore, J.E., Lowery, C.J., Chalmers, R.M., Sulaiman, I., Elwin, K., Rooney, P.J., Millar, B.C., Dooley, J.S., Lal, A.A., Xiao, L., 2002. Three drinking-water associated Cryptosporidiosis outbreaks, Northern Ireland. Emerg. Infect. Dis. 8, 631-633.

\bibitem{grassly} Grassly, N.C., Fraser, C., 2006. Seasonal infectious disease epidemiology. Proc. Roy. Soc. B. 273, 2541-2550.

\bibitem{harper} Harper, G.J., 1961. Airborne micro-organisms: survival tests with four viruses. J. Hyg. (Lond). 59, 479-486.

\bibitem{hemmes} Hemmes, J.H., Winkler, K.C., Kool, S.M., 1960. Virus survival as a seasonal factor in Influenza and Poliomyelitis. Nature. 188, 430-431.

\bibitem{joh} Joh, R.I., Wang, H., Weiss, H., Weitz, J.S., 2009. Dynamics of indirectly transmitted infectious diseasees with immunological threshold. Bull. Math. Biol. 71, 845-862.

\bibitem{keeling} Keeling, M.J., Rohani, P., Grenfell, B.T., 2001. Seasonally forced disease dynamics explored as switching between attractors. Physica D. 148, 317-335.

\bibitem{li} Li, S., Eisenberg, J.N.S., Spicknall, I.H., Koopman, J.S., 2009. Dynamics and control of infections transmitted from person to person through the environment. Am. J. Epidemiol. 170, 257-265.

\bibitem{lofgren} Lofgren, E., Fefferman, N.H., Naumov, Y.N., Gorski, J., Naumova, E.N., 2007. Influenza seasonality: underlying causes and modeling theories. J. Virol. 81, 5429-5436.

\bibitem{lowen} Lowen, A.C., Mubareka, A., Steel, J., Palese, P., 2007. Influenza virus transmission is dependent on relative humidity and temperature. PLoS Pathogens. 3, 1470-1476.

\bibitem{who} Medema, G., Teunis, P., Blokker, M., Deere, D., Charles, P., Loret, J.F., 2009. Risk assessment of Cryptosporidium in drinking water. World Health Organization.

\bibitem{miller} Miller, M.W., Thompson Hobbs, N., Tavener, S.J., 2006. Dynamics of prion disease transmission in mule deer. Ecological Applications. 16, 2208-2214.

\bibitem{nelson} Nelson, R.J., Demas, G.E., 1996. Seasonal changes in immune function. Quart. Rev. Biol. 71, 511-548.

\bibitem{nicas} Nicas, M., Nazaroff, W.W., Hubbard, A., 2005. Towards understanding the risk of secondary airborne infection: emission of respirable pathogens. J. Occup. Environ. Hyg. 2, 143-154.

\bibitem{nieuwhof} Nieuwhof, G.J., Conington, J., Bishop, S.C., 2009. A genetic epidemiological model to describe resistance to an endemic bacterial disease in livestock: application to footrot in sheep. Genetics Selection Evolution. 41, 19.

\bibitem{peeters} Peeters, J.E., Maz\'{a}s, E.A., Masschelein, W.J., Villacorta Martiez de Maturana, I., Debacker, E., 1989. Effect of disinfection of drinking water with ozone or chlorine dioxide on the survival of Cryptosporidium parvum oocysts. Appl. Environ. Microbiol. 55, 1519-1522.

\bibitem{reluga} Reluga, T., 2004. A two-phase epidemic driven by diffusion. J. Theor. Biol. 229, 249-261.

\bibitem{robinson} Robinson, M., Stilianakis, N.I., Drossinos, Y., 2012. Spatial dynamics of airborne infectious diseases. J. Theor. Biol. 297, 116-126.

\bibitem{schulman} Schulman, J.L., Kilbourne, E.D., 1963. Experimental transmission of Influenza virus infection in mice II: some factors affecting the incidence of transmitted infection. J. Exp. Med. 118, 267-275.

\bibitem{shaman09} Shaman, J., Kohn, M., 2009. Absolute humidity modulates influenza survival, transmission, and seasonality. PNAS. 106, 3243-3248.

\bibitem{shaman2010} Shaman, J., Pitzer, V.E., Viboud, C., Grenfell, B.T., Lipsitch, M., 2010. Absolute humidity and the seasonal onset of influenza in the continential United States. PLoS Biology. 8, e10000316.

\bibitem{stilianakis} Stilianakis, N.I., Drossinos, Y., 2010. Dynamics of infectious disease transmission by inhalable respiratory droplets. J. Roy. Soc. Interface. 50, 1355-1366.

\bibitem{stohr} Stohr, K., 2002. Influenza - WHO cares. Lancet Infect. Dis. 2, 517.


\bibitem{stone} Stone, L., Olinky, R., Huppert, A., 2007. Seasonal dynamics of recurrent epidemics. Nature. 446, 533-536.

\bibitem{tien} Tien, J.H., Earn, D.J.D., 2010. Multiple transmission pathways and disease dynamics in a waterborne pathogen model. Bull. Math. Biol. 72, 1506-1533.

\bibitem{truscott} Truscott, J., Fraser, C., Cauchemez, S., Meeyai, A., Hinsley, W., Donnelly, C.A., Ghani, A., Ferguson, N., 2012. Essential epidemiological mechanisms underpinning the transmission dynamics of seasonal influenza. J. R. Soc. Interface. 67, 304-312.

\bibitem{weber} Weber, T.P., Stilianakis, N.I., 2008. Inactivation of influenza A viruses in the environment and modes of transmission: a critical review. J. Infect. 57, 361-373.

\bibitem{yang} Yang, Y., Sugimoto, J.D., Halloran, M.E., Basta, N.E., Chao, D.L., Matrajt, L., Potter, G., Kenah, E., Longini Jr, I.M, 2009. The transmissibility and control of pandemic influenza A (H1N1) virus. Science. 326, 729-733.

\end{thebibliography}
\end{document}